\documentclass[reprint,amsmath,amssymb,aps]{revtex4-2}

\usepackage{xcolor,graphicx}        
\usepackage{float}
\usepackage{amsmath,amsfonts,amssymb}
\usepackage{tikz}
\usepackage{braket}
\usepackage{hyperref}
\hypersetup{
	colorlinks = true,
	linkcolor = red,
	anchorcolor = blue,
	citecolor = blue,
	filecolor = blue,
	urlcolor = blue}
\usepackage[top=2cm,bottom=2cm,right=2cm,left=2cm]{geometry}

\begin{document}

\title{Waveguide arrays interaction to second neighbors: Exact solution}

\author{M. A. Tapia-Valerdi}
%\email[e-mail: ]{mvalerdi@inaoep.mx}
\affiliation{Instituto Nacional de Astrofísica Óptica y Electrónica (INAOE)\\ Luis Enrique Erro 1, Santa María Tonantzintla, Puebla, 72840, Mexico}
\author{I. Ramos-Prieto}
\email[e-mail: ]{iran@inaoep.mx}
\affiliation{Instituto Nacional de Astrofísica Óptica y Electrónica (INAOE)\\ Luis Enrique Erro 1, Santa María Tonantzintla, Puebla, 72840, Mexico}
\author{F. Soto-Eguibar}
%\email[e-mail: ]{feguibar@inaoep.mx}
\affiliation{Instituto Nacional de Astrofísica Óptica y Electrónica (INAOE)\\ Luis Enrique Erro 1, Santa María Tonantzintla, Puebla, 72840, Mexico}
\author{H. M. Moya-Cessa}
%\email[e-mail: ]{hmmc@inaoep.mx}
\affiliation{Instituto Nacional de Astrofísica Óptica y Electrónica (INAOE)\\ Luis Enrique Erro 1, Santa María Tonantzintla, Puebla, 72840, Mexico}

\begin{abstract}
We provide an analytical framework for describing the propagation of light in waveguide arrays, considering both infinite and semi-infinite cases. The interaction up to second neighbors is taken into account, which makes for a more realistic setup. We show that these solutions follow a distinctive structural pattern. This pattern reflects a transition from conventional Bessel functions to the lesser-known one-parameter generalized Bessel functions, offering new insights into the propagation dynamics in these systems.
\end{abstract}

\date{\today}

\maketitle

\section{Introduction}  
Discreteness is a fundamental concept prevalent in many areas of modern physics, such as quantum mechanics. However, the experimental verification of theoretical predictions regarding discrete systems is often challenging. Consequently, systems that exhibit analogous structures in both their physical behavior and mathematical formalism are needed. Arrays of evanescently coupled waveguides serve as a prominent example of such systems \cite{christodoulides2003discretizing,barral2020quantum}. In quantum optics, the ability to design and control the dynamics of discrete coupling or tunneling processes between periodically arranged potential wells is crucial. This topic has gained significant importance in scientific research, and arrays of weakly coupled waveguides have been used to investigate a wide range of phenomena, including quantum walks, Anderson localization, the simulation of $\mathcal{PT}$ symmetric Hamiltonians, and quantum state generation, among other applications \cite{perets2008realization,peruzzo2010quantum,lahini2008anderson,joglekar2013optical,rai2010quantum,perez2016generalized,biggerstaff2016enhancing,paulisch2016universal,Perez-Leija:10,Ancheyta_2017,PhysRevLett.107.103601,Ramos_2021,Leonardi_2023,Urzua_2024}.

The development of new schemes to extend the concept of waveguides in other settings is of significant importance. In many of these structures, only couplings to the nearest-neighboring waveguides are considered \cite{perez2011,schafer2020tools,vicencio2015observation}, while interactions with more distant waveguides are often neglected due to the assumption that the coupling strength decays exponentially with distance, rendering second-order couplings negligible. However, under certain conditions, the influence of a second interaction becomes remarkably significant. For example, in quantum computation and quantum information processing using optical waveguides, it is crucial to fabricate compact waveguide circuits to minimize their footprints \cite{meany2015laser}. As the separation between waveguides in these circuits decreases or as the waveguide length increases, higher-order couplings must be considered. The two-dimensional zigzag waveguide lattice, which is topologically equivalent to a one-dimensional waveguide lattice, has been used to study second-order couplings \cite{efremidis2002discrete,szameit2008long}. The benefits of non-nearest-neighbor couplings have been well documented in the literature, such as modeling quantum states, Bloch oscillations, and photon-number correlations\cite{villegas2022modeling,wang2010nontrivial,dreisow2011observation,qi2014photon}. Moreover, the role of boundaries and surfaces introduce additional complexities to wave dynamics by breaking translational symmetry, significantly altering the interaction landscape that is not present in infinite arrays \cite{longhi2006tunneling,istrate2005photonic}. Experimentally, this can be managed by altering the separation between the boundary waveguide and its neighbors or by adjusting its refractive index, which in turn affects the periodicity of the entire array \cite{ablowitz2003discrete,trompeter2003tailoring}. These modifications affect both nearest-neighbor and nonnearest-neighbor interactions, underscoring the nuanced relationship between boundary effects and waveguide couplings.

 The article is organized as follows. In Section \ref{first infinite}, we start by analyzing the fully infinite waveguide array in the case where only first-neighbor interactions are allowed. In Section \ref{first neighbors}, we expand on this by studding the semi-infinite waveguide array, demonstrating that the system's evolution operator can be factorized in terms of Bessel functions. In Section \ref{second neighbor}, we extend the analysis considering second-neighbor interactions in the complete infinity case. Using the generating function for one-parameter generalized Bessel functions, we derive the optical field, revealing a correlation between the solutions for the semi-infinite and infinite cases, as well as for first- and second-neighbor interactions. In Section \ref{second semi-infinite}, we propose a factorization for the evolution operator in the semi-infinite case with second-neighbor interactions, and applying the method of Section \ref{first neighbors}, we demonstrate its validity. We then calculate the evolution of the amplitude for each waveguide and apply this solution to standard initial conditions in quantum optics. The article concludes with Section \ref{conclusiones}, where we present our final remarks.

\section{Interaction to first neighbor's: Case in which the operators are infinite on both sides}\label{first infinite}
In coupled mode theory, the propagation of an optical field through a waveguide array with nearest-neighbor evanescent coupling without frontiers is governed by the following set of coupled first order ordinary differential equations
\begin{equation}\label{dif_eq1}
i\frac{dE_j(z)}{dz}= g_1(E_{j-1}+E_{j+1}), \qquad j=-\infty,\dots,\infty,
\end{equation}
with $z$ the propagation distance, $g_1$ the coupling constant and $j$ runs from $-\infty$ to $\infty$ through all the integers.\\
This framework can be viewed also as the problem of solving the Schrödinger-like equation with Hamiltonian
\begin{equation}\label{0020}
\hat{H}=g_1\left( \hat{V}_{\infty}+\hat{V}_{\infty}^\dagger\right),
\end{equation}
where the operators $\hat{V}_{\infty}$ and $\hat{V}_{\infty}^\dagger$ are the infinite counterparts of the London-Susskind–Glogower operators \cite{london1926jacobischen,susskind1964quantum}, defined as
\begin{equation}
\hat{V}_{\infty}= \sum_{n=-\infty}^{\infty}\ket{n}\bra{n+1}, \quad \hat{V}_{\infty}^\dagger= \sum_{n=-\infty}^{\infty}\ket{n+1}\bra{n};
\end{equation}
the states $\ket{n}, \; n \in \mathbb{Z}$, are generalized Fock states, and it is possible to show that $\left[\hat{V}_{\infty},\hat{V}_{\infty}^\dagger\right]=0$ as well as $\hat{V}_{\infty}\hat{V}_{\infty}^\dagger=\hat{I}$.\\
If we denote the solution of the Schrödinger-type equation obtained from the Hamiltonian \eqref{0020} as $\ket{\psi(z)}$, it is easy to demonstrate that the solutions of system \eqref{dif_eq1} are given by $E_j(z)=\braket{j|\psi(z)}, \; j \in \mathbb{Z}$.\\
Using operational methods, we proceed to solve the Schrödinger-like equation obtained from the Hamiltonian \eqref{0020}, and subsequently derive the solution for the infinite system in \eqref{dif_eq1}. Given the initial condition $\ket{\psi(0)}=\ket{n_0}$, the formal solution of the Schrödinger-like equation can be written as $\ket{\psi_{n_0}(z)}=\exp\left[-i g_1 z  \left(\hat{V}_{\infty}+\hat{V}_{\infty}^\dagger\right)\right]\ket{n_0}$; to calculate this expression, we use that $\hat{V}_{\infty}\hat{V}_{\infty}^\dagger=\hat{I}$ implies $\hat{V}_{\infty}=\frac{1}{\hat{V}_{\infty}^\dagger}$; hence, we can write 
\begin{equation}
\ket{\psi_{n_0}(z)}=\exp\left[-ig_1 z\left(i\hat{V}_{\infty}^\dagger-\frac{1}{i\hat{V}_{\infty}^\dagger}\right)\right]\ket{n_0}. 
\end{equation}
Since the generating function for the Bessel functions of the first kind \cite{arfken2011mathematical} is apparent in the preceding equation, we can reformulate the solution to the Schrödinger-like equation as
\begin{equation}
\ket{\psi_{n_0}(z)}=\sum_{n=-\infty}^{\infty}J_n(-2g_1z)i^n \hat{V}_{\infty}^{\dagger n}\ket{n_0}.
\end{equation}
Using that $\hat{V}_{\infty}^{\dagger n}\ket{n_0}=\ket{n_0+n}$, we get
\begin{equation}
\ket{\psi_{n_0}(z)}=\sum_{n=-\infty}^{\infty}J_n(-2g_1z)i^n \ket{n_0+n},
\end{equation}
and in order to find the amplitudes $E_j$ that are the solutions of the system \eqref{dif_eq1}, we simply multiply by the bra $\bra{j}$ and obtain
\begin{equation}\label{0070}
E_{n_0,j}(z)=\braket{j|\psi_{n_0}(z)}=i^{j-n_0}  J_{j-n_0}(-2g_1 z), \quad j \in \mathbb{Z}.
\end{equation}

\section{Interaction to first neighbor's: Case semi-infinite }\label{first neighbors}
In the case of propagation of an optical field through a waveguide array with nearest-neighbor evanescent coupling with frontiers the set of coupled differential equations which rules the behavior is
\begin{align}
i\frac{dE_0(z)}{dz} &= g_1 E_1,
\nonumber \\
i\frac{dE_j(z)}{dz} &= g_1 (E_{j-1}+E_{j+1}), \quad j=1,2,3,\dots,
\end{align}
being $g_1$ again the coupling constant.\\
As in the previous case, this system of equations is equivalent to the Schrödinger-like equation $i\frac{d\ket{\psi(z)}}{dz}=\hat{H}\ket{\psi(z)}$ considering the Hamiltonian $\hat{H}=g_1 \left(\hat{V}+\hat{V}^\dagger\right)$, where $\hat{V}$ and $\hat{V}^{\dagger}$ are the known London-Susskind-Glogower operators \cite{london1926jacobischen,susskind1964quantum} and $\ket{\psi(z)} = \sum_{j=0}^\infty E_j(z)\ket{j}$. The evolution operator for our Hamiltonian can then be expressed as $\hat{U}(z)=\exp[-ig_1z(\hat{V}+ \hat{V}^{\dagger})]$; by employing operational methods \cite{doi:10.1142/S0219749911007319}, it is possible to show that it can be rewritten as
\begin{align}\label{op-evol12}
\hat{U}(z)=&\sum_{n,m=0}^{\infty}\left[i^{m-n}J_{m-n}(-2g_1z)\right.
\nonumber \\
& \left.+i^{n+m}J_{m+n+2}(-2g_1z)\right]\hat{V}^{\dagger m}\ket{0}\bra{0}\hat{V}^n.
\end{align}
In order to demonstrate the method that will later be used to verify the correctness of our ansatz for second-neighbor interactions in the semi-infinite case, we will first prove the validity of Eq.~\eqref{op-evol12}. We start by differentiating both sides of the equation with respect to $z$, then group terms and separate one of the sums over $m$. On the one hand, this leads to the expression
\begin{align}\label{0420}
\frac{d\hat{U}(z)}{dz}=&\sum_{n,m=0}^{\infty}g_1\left[i^{m-n}J_{m-n+1}(-2g_1z)\right.
\nonumber \\
& \left.+i^{n+m}J_{m+n+3}(-2g_1z) \right]\hat{V}^{\dagger m}|0\rangle\langle 0|\hat{V}^n
\nonumber \\ 
&-\sum_{n=0,m=1}^{\infty}g_1\left[i^{m-n}J_{m-n-1}(-2g_1z)\right.
\nonumber \\
& \left. +i^{n+m}J_{m+n+1}(-2g_1z)\right]\hat{V}^{\dagger m}|0\rangle\langle 0|\hat{V}^n,
\end{align}
where we have used the Bessel functions property $J_{-n}(z)=(-1)^n J_n(z), n \in \mathbb{Z}$ that implies $i^{-n} J_{-n-1}(-2g_1z) + i^n J_{n+1}(-2g_1z) = 0$, which we used to simplify the result.

Additionally, based on Eq.~(\ref{op-evol12}) and keeping in mind that $\hat{U}(z)=\exp[-i g_1 z(\hat{V}+\hat{V}^{\dagger})]$, we can derive that
\begin{align}
\frac{d\hat{U}(z)}{dz}=&-i g_1 (\hat{V}+\hat{V}^{\dagger})\sum_{n=0,k=0}^{\infty}\left[i^{k-n}J_{k-n}(-2g_1z)\right.
\nonumber \\
& \left. +i^{n+k}J_{k+n+2}(-2g_1z)\right]\hat{V}^{\dagger k}|0\rangle\langle 0|\hat{V}^n.
\end{align}
It is possible to show from the commutation relations between $V$ and $V^\dagger$ that $[\hat{V},\hat{V}^{\dagger k}]=\hat{V}^{\dagger (k-1)}|0\rangle\langle 0|$.
Performing the product, and using the previous equation, it is easy to see that the derivative can be written as
\begin{align} \label{0460}
\frac{d\hat{U}(z)}{dz}=&-i g_1\left\{\sum_{n,m=0}^{\infty}\left[i^{m-n+1}J_{m-n+1}(-2g_1z)\right.\right.
\nonumber \\
& \quad \left.\left.+i^{n+m+1}J_{m+n+3}(-2g_1z) \right]\hat{V}^{\dagger m}|0\rangle\langle 0|\hat{V}^n\right\}
\nonumber \\ 
&-i g_1 \left\{\sum_{n=0,m=1}^{\infty}\left[i^{m-n-1}J_{m-n-1}(-2g_1z)\right.\right.
\nonumber \\
& \quad \left.\left. +i^{n+m-1}J_{m+n+1}(-2g_1z)\right]\hat{V}^{\dagger m}|0\rangle\langle 0|\hat{V}^n\right\},
\end{align}
where we have made the index substitutions $m=k-1$ and $m=k+1$ when needed; as was anticipated, Eqs.~\eqref{0420} and \eqref{0460} are exactly equivalent.\\
It is evident that the optical field at the $j$-th site, when the $n_0$-th guide is excited, is given by
\begin{align}
E_{n_0,j}(z)=&i^{j-n_0}J_{j-n_0}(-2g_1z)+i^{j+n_0}J_{j+n_0+2}(-2g_1z),
\nonumber \\
&\quad j=0,1,2,\dots,\infty.
\end{align}
This result aligns with the work of Makris and Christodoulides \cite{makris2006method}, who employed the method of images to analyze a finite one-dimensional array of $N$ waveguides.

In the next sections, we show how to extend the solution for the interaction to the second neighbors. Although the solution to first neighbors is difficult to obtain, we show that it may be generalized by using extensions of the Bessel functions introduced by Dattoli {\it et al.} \cite{dattoli1991theory}.

\section{Interaction to second neighbor's: Case in which the operators are infinite on both sides}\label{second neighbor}
Under coupled mode theory, the dynamics of an optical field in a waveguide array with next-nearest-neighbor evanescent coupling is described by the following set of coupled differential equations 
\begin{equation}\label{0140}
i\frac{dE_j(z)}{dz} = g_1(E_{j-1}+E_{j+1})+g_2(E_{j-2}+E_{j+2}),
\end{equation}
with $j$ running through all the integers $\mathbb{Z}$.\\
As in the previous sections, this system is equivalent to a Schrödinger-like equation with the Hamiltonian $\hat{H}=g_1\left( \hat{V}_{\infty}+\hat{V}_{\infty}^\dagger\right)  + g_2\left(\hat{V}_{\infty}^2+\hat{V}_{\infty}^{\dagger2}\right)$. The evolution operator corresponding to this Hamiltonian is
\begin{equation}\label{0050}
\hat{U}(z)=\exp \left\{  -iz \left[g_1 \left(\hat{V}_{\infty}+\hat{V}_{\infty}^\dagger\right)+g_2 \left(\hat{V}_{\infty}^2+\hat{V}_{\infty}^{\dagger 2}\right)\right] \right\}.
\end{equation}
Using again that $\hat{V}_{\infty}=\frac{1}{\hat{V}_{\infty}^\dagger}$, we can cast this evolution operator as
\begin{equation}\label{0160}
\hat{U}(z)=\exp \left[-g_1 z \left(i\hat{V}_{\infty}^\dagger-\frac{1}{i\hat{V}_{\infty}^\dagger}\right)-g_2 z \left(i\hat{V}_{\infty}^{\dagger 2}-\frac{1}{i\hat{V}_{\infty}^{\dagger 2}}\right)\right].
\end{equation}
It is convenient to introduce the not so well known one parameter generalized Bessel functions $J_n(x,y;s)$ \cite{dattoli1991theory} defined as 
\begin{equation}
J_n(x,y;s)=\sum_{k=-\infty}^{\infty}s^k J_{n-2k}(x) J_k(y),
\end{equation}
whose generating function is 
\begin{equation}\label{0080}
\sum _{n=-\infty }^{\infty } t^n J_n(x,y;s)=
\exp\left[\frac{x}{2} \left(t-\frac{1}{t}\right) + \frac{y}{2}  \left(st^2-\frac{1}{st^2}\right)\right].
\end{equation}
The two operators in the propagator Eq.~\eqref{0160},  $i\hat{V}_{\infty}^\dagger-\frac{1}{i\hat{V}_{\infty}^\dagger}$ and $i\hat{V}_{\infty}^{\dagger 2}-\frac{1}{i\hat{V}_{\infty}^{\dagger 2}}$ commute, and we can identify the terms with those of the generating function \eqref{0080}; thus, if in the generating function Eq.~\eqref{0080} we identify $t\rightarrow i\hat{V}_{\infty}^\dagger$, $x\rightarrow -2g_1z$, $y\rightarrow -2g_2z$, $s\rightarrow -i$, we obtain
\begin{equation}
\hat{U}(z)=\sum _{n=-\infty }^{\infty } \left( i\hat{V}_{\infty}^{\dagger}\right)^n J_n(-2g_1z,-2g_2z;-i).
\end{equation}
Rearranging and commuting,
\begin{equation}
\hat{U}(z)=\sum _{n=-\infty }^{\infty } i^n  J_n(-2g_1z,-2g_2z;-i) \hat{V}_\infty^{\dagger n},
\end{equation}
and we have the solution to our problem.\\
If we have an initial condition $\ket{\psi(0)}$, we get
\begin{equation}
\ket{\psi(z)}=\sum _{n=-\infty }^{\infty } i^n  J_n(-2g_1z,-2g_2z;-i) \hat{V}_\infty^{\dagger n}\ket{\psi(0)}.
\end{equation}
Let us specify the initial condition setting $\ket{\psi(0)}=\ket{n_0}$, with $n_0$ an integer; as $\hat{V}_\infty^{\dagger n}\ket{n_0}=\ket{n_0+n}$,
\begin{equation}\label{0130}
\ket{\psi_{n_0}(z)}=\sum _{n=-\infty }^{\infty } i^n  J_n(-2g_1z,-2g_2z;-i)\ket{n_0+n},
\end{equation}
and so we have
\begin{align}\label{0230}
E_{n_0,j}(z)=&\braket{j|\psi_{n_0}(z)}=i^{j-n_0}  J_{j-n_0}(-2g_1 z,-2g_2 z;-i),
\nonumber \\
j=&-\infty, \dots,-2,-1,0,1,2,\dots,\infty.
\end{align}
The structure of this result is similar to that of the infinite case of an optical field propagating through a waveguide array with nearest-neighbor evanescent coupling, Eq. \eqref{0070}, except that the Bessel functions are replaced by the one parameter generalized Bessel functions. It is easy to show that for $z=0$ the solution \eqref{0230} reduces to the initial condition $E_{n_0,j}(0)=\delta_{n_0,j}$, and by direct substitution of the said solution in the system of equations \eqref{0140} it is found that it is indeed the solution.

\begin{figure}[h!]
\includegraphics{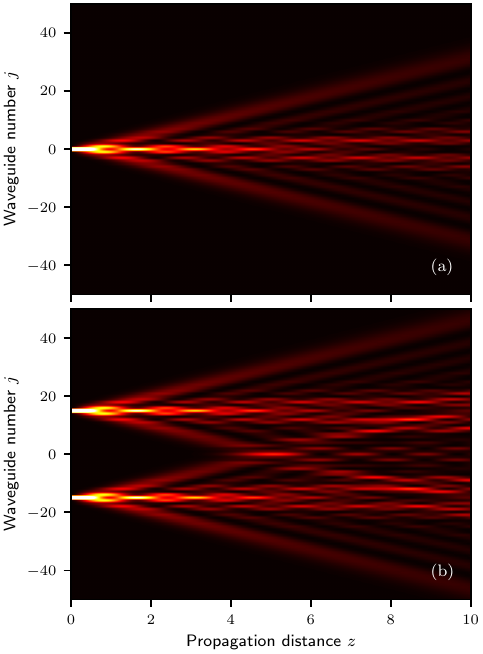}
\caption{The evolution of the squared amplitude modulus in each waveguide is depicted for an infinite waveguide array governed by Eq.~\eqref{0230}, with the parameters $g_1=1.0$ and $g_2=0.5$; (a) The central guide, $j=0$ is illuminated, (b) The guides $j=-15$ and $j=15$ are illuminated.}
\label{fig1}
\end{figure}
In Fig.~\ref{fig1}, we present the intensity distribution in the waveguides as a function of the propagation distance $z$ given by \eqref{0230}. We assume $g_1 > g_2$ to be a realistic condition, based on the fact that the coupling strength between the waveguides decreases as the light propagates farther along the array. In panel (a), the central guide is illuminated. To show the interaction between the guides, in panel (b), two guides are excited, $j=-15$ and $j=15$.

\section{Interaction to second neighbor's: Case semi-infinite}\label{second semi-infinite}
In the following, we will analyze the dynamics of an optical field in a waveguide array with next-nearest-neighbor evanescent coupling with boundaries. The system is described by the following coupled differential equations
\begin{align}\label{0240}
i\frac{dE_0(z)}{dz} &=g_1 E_1 + g_2(E_2-E_0), \nonumber \\
i\frac{dE_1(z)}{dz} &=g_1(E_0+E_2) + g_2E_3, \\
i\frac{dE_j(z)}{dz} &=g_1(E_{j-1}+E_{j+1})+g_2(E_{j-2}+E_{j+2}), \nonumber
\end{align}
where $j=2,3,4,\dots$.\\
Now, we consider the Hamiltonian $\hat{H}=g_1(\hat{V}+\hat{V}^\dagger)+g_2(\hat{V}^2+\hat{V}^{\dagger 2}-\ket{0}\bra{0})$, where once again $\hat{V}$ and $\hat{V}^{\dagger}$ are the usual Susskind–Glogower operators \cite{london1926jacobischen,susskind1964quantum}, and the evolution operator may be written as
\begin{equation}\label{op-evol}
\hat{U}(z)=\exp\{-iz[g_1(\hat{V}+ \hat{V}^{\dagger}) + g_2(\hat{V}^2+\hat{V}^{\dagger 2}- \ket{0}\bra{0})]\}.
\end{equation}
We propose an ansatz for the evolution operator, maintaining the same structure as Eq.~\eqref{op-evol12}, but replacing the Bessel functions with one-parameter generalized Bessel functions \cite{dattoli1991theory}:
\begin{align}\label{op-evol2}
\hat{U}(z)=&\sum_{n,m=0}^{\infty}\left[i^{m-n}J_{m-n}(-2g_1z,-2g_2z;-i)\right.
\nonumber \\
&\left.+i^{n+m}J_{m+n+2}(-2g_1z,-2g_2z;-i)\right]\hat{V}^{\dagger m}|0\rangle\langle 0|\hat{V}^n.
\end{align}
To demonstrate the validity of our ansatz, as done in Section \ref{first neighbors}, we differentiate the above expression with respect to $z$. In Appendix \ref{appendix 1}, we show that this yields the result
\begin{align}\label{solution}
&\sum_{n=0,m=0}^{\infty} g_1\left[i^{m-n}J_{m-n+1}(-2g_1z,-2g_2z;-i)\right.
\nonumber \\
& \left. +i^{n+m}J_{m+n+3}(-2g_1z,-2g_2z;-i)\right]\hat{V}^{\dagger m}\ket{0}\bra{0}\hat{V}^n
\nonumber \\
& -\sum_{n=0,m=1}^{\infty}g_1\left[i^{m-n}J_{m-n-1}(-2g_1z,-2g_2z;-i)\right.
\nonumber \\
& \left. +i^{n+m}J_{m+n+1}(-2g_1z,-2g_2z;-i)\right]\hat{V}^{\dagger m}\ket{0}\bra{0}\hat{V}^n
\nonumber \\  
&+ \sum_{n=0,m=0}^{\infty}g_2\left[i^{m-n+1}J_{m-n+2}(-2g_1z,-2g_2z;-i)\right.
\nonumber \\
& \left. +i^{n+m+1}J_{m+n+4}(-2g_1z,-2g_2z;-i)\right]\hat{V}^{\dagger m}\ket{0}\bra{0}\hat{V}^n
\nonumber \\
&-\sum_{n=0,m=2}^{\infty}g_2\left[i^{m-n-1}J_{m-n-2}(-2g_1z,-2g_2z;-i)\right.
\nonumber \\
& \left. +i^{n+m-1}J_{m+n}(-2g_1z,-2g_2z;-i)\right]\hat{V}^{\dagger m}\ket{0}\bra{0}\hat{V}^n
\nonumber \\
& -\sum_{n=0}^{\infty}g_2\left[i^{n-1}J_{n+2}(-2g_1z,-2g_2z;-i)\right.
\nonumber \\
& \left.+i^{n-1}J_n(-2g_1z,-2g_2z;-i)\right]\ket{0}\bra{0}\hat{V}^n.
\end{align}
On the other hand, from  equations (\ref{op-evol}) and (\ref{op-evol2}), we have that
\begin{align}
\frac{d\hat{U}(z)}{dz}=&-i[g_1(\hat{V}+\hat{V}^{\dagger})+g_2(\hat{V}^2+\hat{V}^{\dagger 2}-\ket{0}\bra{0})]
\nonumber \\
&\sum_{n,k=0}^{\infty}\left[i^{k-n}J_{k-n}(-2g_1z,-2g_2z;-i)\right.
\nonumber \\
& \left. +i^{n+k}J_{k+n+2}(-2g_1z,-2g_2z;-i)\right]\hat{V}^{\dagger k}|0\rangle\langle 0|\hat{V}^n;
\end{align}
calculating the product of the operators, recognizing that $\hat{V}^k\ket{0} = \ket{k}$, and applying the commutation rules, we obtain
\begin{align}
&-i g_1\sum_{n=0,k=1}^{\infty}\left[i^{k-n}J_{k-n}(-2g_1z,-2g_2z;-i)\right.
\nonumber \\
& \left. +i^{n+k}J_{k+n+2}(-2g_1z,-2g_2z;-i)\right]\hat{V}^{\dagger(k -1)}\ket{0}\bra{0}\hat{V}^n 
\nonumber \\
& -ig_1 \sum_{n=0,k=0}^{\infty}\left[i^{k-n}J_{k-n}(-2g_1z,-2g_2z;-i)\right.
\nonumber \\
& \left. +i^{n+k}J_{k+n+2}(-2g_1z,-2g_2z;-i)\right]\hat{V}^{\dagger(k +1)}\ket{0}\bra{0}\hat{V}^n 
\nonumber \\
& -ig_2 \sum_{n=0,k=2}^{\infty}\left[i^{k-n}J_{k-n}(-2g_1z,-2g_2z;-i)\right.
\nonumber \\
& \left. +i^{n+k}J_{k+n+2}(-2g_1z,-2g_2z;-i)\right]\hat{V}^{\dagger(k -2)}\ket{0}\bra{0}\hat{V}^n
\nonumber \\
& -ig_2 \sum_{n=0,k=0}^{\infty}\left[i^{k-n}J_{k-n}(-2g_1z,-2g_2z;-i)\right.
\nonumber \\
& \left.+i^{n+k}J_{k+n+2}(-2g_1z,-2g_2z;-i)\right]\hat{V}^{\dagger(k +2)}\ket{0}\bra{0}\hat{V}^n
\nonumber \\
& +ig_2 \sum_{n=0}^{\infty}\left[i^{-n}J_{-n}(-2g_1z,-2g_2z;-i)\right.
\nonumber \\
& \left.+i^{n}J_{n+2}(-2g_1z,-2g_2z;-i)\right]\ket{0}\bra{0}\hat{V}^n. 
\end{align}
Assigning $m = k - 1$ in the first sum, $m = k + 1$ in the second, $m = k - 2$ in the third, $m = k + 2$ in the fourth, and applying  Eq.~\eqref{recurence}, we obtain the final expression
\begin{align}\label{opevo2}
& -ig_1 \sum_{n=0,m=0}^{\infty}\left[i^{m-n+1}J_{m-n+1}(-2g_1z,-2g_2z;-i)\right.
\nonumber \\
& \left. +i^{m+n+1}J_{m+n+3}(-2g_1z,-2g_2z;-i)\right]\hat{V}^{\dagger m}\ket{0}\bra{0}\hat{V}^n 
\nonumber \\
& -ig_1 \sum_{n=0,m=1}^{\infty}\left[i^{m-n-1}J_{m-n-1}(-2g_1z,-2g_2z;-i)\right.
\nonumber \\
&\left. +i^{m+n-1}J_{m+n+1}(-2g_1z,-2g_2z;-i)\right]\hat{V}^{\dagger m}\ket{0}\bra{0}\hat{V}^n 
\nonumber \\
& -ig_2 \sum_{n=0,m=0}^{\infty}\left[i^{m-n+2}J_{m-n+2}(-2g_1z,-2g_2z;-i)\right.
\nonumber \\
& \left. +i^{m+n+2}J_{m+n+4}(-2g_1z,-2g_2z;-i)\right]\hat{V}^{\dagger m}\ket{0}\bra{0}\hat{V}^n
\nonumber \\
& -ig_2 \sum_{n=0,m=2}^{\infty}\left[i^{m-n-2}J_{m-n-2}(-2g_1z,-2g_2z;-i)\right.
\nonumber \\
&\left. +i^{m+n-2}J_{m+n}(-2g_1z,-2g_2z;-i)\right]\hat{V}^{\dagger m}\ket{0}\bra{0}\hat{V}^n
\nonumber \\
&-g_2\sum_{n=0}^{\infty}\left[i^{n-1}J_{n+2}(-2g_1z,-2g_2z;-i)\right.
\nonumber \\
&\left.+i^{n-1}J_{n}(-2g_1z,-2g_2z;-i)\right]\ket{0}\bra{0}\hat{V}^n,
\end{align}
When we compare Eqs.~\eqref{solution} and \eqref{opevo2}, the equality is evident, as intended. Therefore, we can express the solution in the form
\begin{align}\label{ketpsi2}
\ket{\psi(z)}=&\sum_{n=0,m=0}^{\infty}\left[i^{m-n}J_{m-n}(-2g_1z,-2g_2z;-i)\right.
\nonumber \\
& +i^{n+m}J_{m+n+2}(-2g_1z,-2g_2z;-i)
\nonumber \\
& \left.\right]\hat{V}^{\dagger m}|0\rangle\langle 0|\hat{V}^n \ket{\psi(0)}.
\end{align}
We consider as initial conditions a Fock state and a coherent state. First, taking $\ket{\psi(0)}=\ket{n_0}$, we get 
\begin{align}
\ket{\psi_{n_0}(z)}=&\sum_{n=0,m=0}^{\infty}\left[i^{m-n}J_{m-n}(-2g_1z,-2g_2z;-i)\right.
\nonumber \\
&\left. +i^{n+m}J_{m+n+2}(-2g_1z,-2g_2z;-i)\right]\ket{m} \braket{n|n_0}.
\end{align}
To obtain the evolution of the amplitude of each guide along the direction propagation, we multiply by the bra $\bra{j}$ to obtain
\begin{align}\label{psi2}
E_{n_0,j}(z)=&i^{j-n_0}J_{j-n_0}(-2g_1z,-2g_2z;-i)
\nonumber \\
& +i^{n_0+j}J_{n_0+j+2}(-2g_1z,-2g_2z;-i),
\\
j=&0,1,2,\dots. \nonumber
\end{align}
As a check, we can set $z=0$ in the solution above, and we get the correct initial condition, i.e., initially only the guide $n_0$ is illuminated; furthermore, if we substitute the solution \eqref{psi2} into the system of equations \eqref{0240}, we verify that it is indeed the solution.

\begin{figure}[h!]
\includegraphics{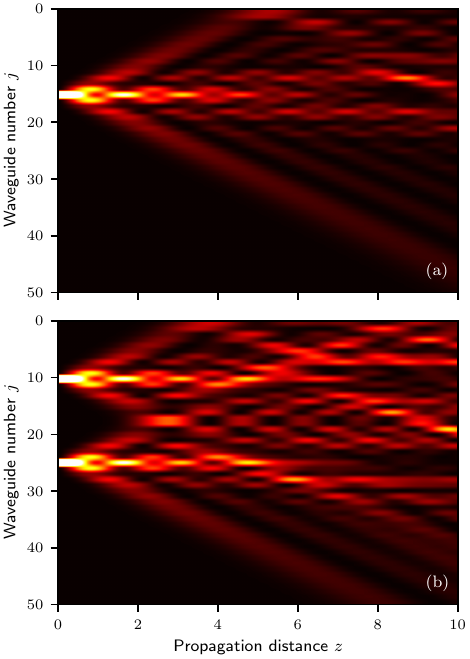}
\caption{Intensity of the field in each waveguide for different initial condition Eq.~\eqref{psi2}; the parameters are $g_1=1.0$ and $g_2=0.5$; (a) The guide $j=15$ is illuminated, (b) The guides $j=10$ and $j=25$ are excited.}
\label{fig2}
\end{figure}

In Fig. \ref{fig2}, we present the intensity distribution in the waveguides as a function of the propagation distance $z$, considering the scenario in which light hits different waveguides. As explained in the preceding paragraph of Fig. \ref{fig1}, we assume $g_1>g_2$. In panel (a), the guide $j=15$ is irradiated, the light travels a certain distance before encountering the edge and interacting with the boundary, resulting in a noticeable reflection, and panel (b) shows the case where two guides are illuminated, one located near the edge and the other more in the center.

We consider now the case when the initial condition is a coherent state $\ket{\psi(0)}=\ket{\alpha}=e^{-\frac{|\alpha|^2}{2}}\sum_{l=0}^{\infty}\frac{\alpha^l}{\sqrt{l!}}\ket{l} $, we get 
\begin{align}
&\ket{\psi_\alpha(z)}=e^{-\frac{|\alpha|^2}{2}}\sum_{l=0,m=0}^{\infty}\left[i^{m-l}J_{m-l}(-2g_1z,-2g_2z;-i)\right.
\nonumber \\
& \qquad \left. +i^{l+m}J_{m+l+2}(-2g_1z,-2g_2z;-i)\right]\ket{m}\frac{\alpha^l}{\sqrt{l!}};
\end{align}
now, we multiply by the bra $\bra{j}$ to obtain the evolution of the amplitude of each guide along the direction propagation as 
\begin{align}\label{psi3}
E_{\alpha,j}(z)=&e^{-\frac{|\alpha|^2}{2}}\sum_{l=0}^{\infty}\frac{\alpha^l}{\sqrt{l!}} \left[i^{j-l}J_{j-l}(-2g_1z,-2g_2z;-i)\right.
\nonumber \\
&\left. +i^{l+j}J_{l+j+2}(-2g_1z,-2g_2z;-i)\right].
\end{align}
In this case, it can also be shown by direct substitution that the above solution satisfies the initial conditions and the system of equations \eqref{0240}.
\begin{figure}[h!]
\includegraphics{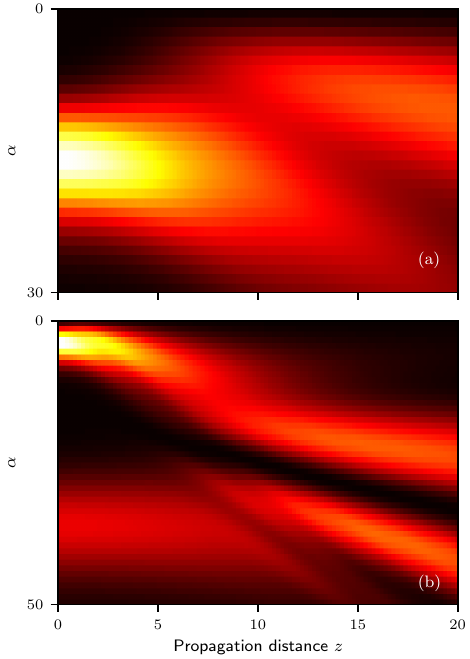}
\caption{Intensity of the field in each waveguide when the initial conditions corresponds to a coherent state, Eq.~\eqref{psi3}. The parameters are $g_1=1.0$ and $g_2=0.5$;(a) Case with $\alpha=4$ (average photon number $\langle\hat{n}\rangle= 16$), (b) Superposition of two coherent states, one with $\alpha=2$ and the other with $\alpha=6$.}
\label{fig3}
\end{figure}
Fig. \ref{fig3} shows the distribution of light intensity along the $z$ axis and across the site number when a coherent light distribution is injected into the initial plane. Similarly to the previous scenario, we assume that it is physically reasonable that $g_1 > g_2$, since the coupling strength between the waveguides diminishes during light transmission from one to the other. This configuration can be achieved using the femtosecond laser writing technique, which allows the creation of large two-dimensional lattices with customizable topologies \cite{szameit2007control,szameit2006hexagonal}. Panel (a) shows the evolution of a coherent state with the average photon number $\braket{\hat{n}}=16$, ($\alpha=4$), illustrating the reflection from the upper boundary of the semi-infinite array. In panel (b), in order to present the interaction between different guides, we introduced two coherent states; the first with a mean photon number of 4 ($\alpha=2$), and the second with an average number of photons of 36 ($\alpha=6$).

\section{Conclusions}\label{conclusiones}
The method for deriving a solution for field propagation in optical lattices, accounting for interaction with second neighbors, has been demonstrated. We have successfully extended the solution for the semi-infinite interaction, initially applied to first-neighbor interactions as described in \cite{makris2006method}. This extension is nontrivial, as the London phase operators do not close to an algebra \cite{doi:10.1142/S0219749911007319}. The complete solution was achieved by adapting fully infinite case solutions, expressed through generalized Bessel functions \cite{dattoli1991theory}, to the semi-infinite context.

\appendix
\section{}\label{appendix 1}
In this appendix, we show that the differentiation of equation \eqref{op-evol2} with respect to $z$ yields the expression given by \eqref{solution}. This result is obtained first applying the recurrence properties of the one-parameter generalized Bessel functions and appropriately rearranging the terms as 
\begin{align}
\frac{d\hat{U}(z)}{dz}=&\sum_{n,m=0}^{\infty} g_1\left[i^{m-n}J_{m-n+1}(-2g_1z,-2g_2z;-i) \right.
\nonumber \\
&\left. +i^{n+m}J_{m+n+3}(-2g_1z,-2g_2z;-i)\right]\hat{V}^{\dagger m}|0\rangle\langle0|\hat{V}^n
\nonumber \\
& -\sum_{n,m=0}^{\infty}g_1\left[i^{m-n}J_{m-n-1}(-2g_1z,-2g_2z;-i) \right.
\nonumber \\
&\left. +i^{n+m}J_{m+n+1}(-2g_1z,-2g_2z;-i)\right]\hat{V}^{\dagger m}|0\rangle\langle0|\hat{V}^n
\nonumber \\
&+ \sum_{n,m=0}^{\infty}g_2\left[i^{m-n+1}J_{m-n+2}(-2g_1z,-2g_2z;-i) \right.
\nonumber \\
& \left. +i^{n+m+1}J_{m+n+4}(-2g_1z,-2g_2z;-i)\right]\hat{V}^{\dagger m}|0\rangle\langle0|\hat{V}^n
\nonumber \\
&-\sum_{n,m=0}^{\infty}g_2\left[i^{m-n-1}J_{m-n-2}(-2g_1z,-2g_2z;-i)\right.
\nonumber \\
&\left. +i^{n+m-1}J_{m+n}(-2g_1z,-2g_2z;-i)\right]\hat{V}^{\dagger m}|0\rangle\langle0|\hat{V}^n,
\end{align}
separating the second and fourth sums with respect to $m$, the previous expression can be rewritten as
\begin{align}\label{a2}
&\sum_{n=0,m=0}^{\infty} g_1\left[i^{m-n}J_{m-n+1}(-2g_1z,-2g_2z;-i)\right.
\nonumber \\
& \left. +i^{n+m}J_{m+n+3}(-2g_1z,-2g_2z;-i)\right]\hat{V}^{\dagger m}\ket{0}\bra{0}\hat{V}^n
\nonumber \\
& -\sum_{n=0,m=1}^{\infty}g_1\left[i^{m-n}J_{m-n-1}(-2g_1z,-2g_2z;-i)\right.
\nonumber \\
& \left. +i^{n+m}J_{m+n+1}(-2g_1z,-2g_2z;-i)\right]\hat{V}^{\dagger m}\ket{0}\bra{0}\hat{V}^n
\nonumber \\
&-\sum_{n=0}^{\infty}g_1\left[i^{-n}J_{-n-1}(-2g_1z,-2g_2z;-i)\right.
\nonumber \\
& \left. +i^{n}J_{n+1}(-2g_1z,-2g_2z;-i)\right]\ket{0}\bra{0}\hat{V}^n
\nonumber \\
&+\sum_{n=0,m=0}^{\infty}g_2\left[i^{m-n+1}J_{m-n+2}(-2g_1z,-2g_2z;-i)\right.
\nonumber \\
& \left. +i^{n+m+1}J_{m+n+4}(-2g_1z,-2g_2z;-i)\right]\hat{V}^{\dagger m}\ket{0}\bra{0}\hat{V}^n
\nonumber \\
&-\sum_{n=0,m=2}^{\infty}g_2\left[i^{m-n-1}J_{m-n-2}(-2g_1z,-2g_2z;-i)\right.
\nonumber \\
& \left. +i^{n+m-1}J_{m+n}(-2g_1z,-2g_2z;-i)\right]\hat{V}^{\dagger m}\ket{0}\bra{0}\hat{V}^n
\nonumber \\
&-\sum_{n=0}^{\infty}g_2\left[i^{-n}J_{-n-1}(-2g_1z,-2g_2z;-i)\right.
\nonumber\\
& \left. +i^{n}J_{n+1}(-2g_1z,-2g_2z;-i)\right]\hat{V}^{\dagger}\ket{0}\bra{0}\hat{V}^n
\nonumber \\
&-g_2\left[i^{-n-1}J_{-n-2}(-2g_1z,-2g_2z;-i)\right.
\nonumber \\
& \left. +i^{n-1}J_n(-2g_1z,-2g_2z;-i)\right]\ket{0}\bra{0}\hat{V}^n.
\end{align}
To simplify these result, we use the following relations for the one parameter generalized Bessel function $J_{-n}(x,y;s)=J_n(-x,-y,1/s)$, $J_{n}(-x,y;s)=(-1)^n J_n(x,y;s)$ as well as $J_{n}(x,-y;s)=J_n(x,y;-s)$, one can be convinced that for the specific case in which $s=-i$ it is true that
\begin{equation}\label{recurence}
 J_{-n}(x,y;-i)=(-1)^n J_n(x,y;-i),  
\end{equation}
so for all non-negative integer $n$, it is easy to show that
\begin{align*}
&\sum_{n=0}^{\infty}g_1\left[i^{-n}J_{-n-1}(-2g_1z,-2g_2z;-i)\right.\\
&\left.+i^{n}J_{n+1}(-2g_1z,-2g_2z;-i)\right]\ket{0}\bra{0}\hat{V}^n=0, \\
&\sum_{n=0}^{\infty}g_2\left[i^{-n}J_{-n-1}(-2g_1z,-2g_2z;-i)\right.\\
&\left.+i^{n}J_{n+1}(-2g_1z,-2g_2z;i)\right]V^{\dagger}\ket{0}\bra{0}\hat{V}^n=0, \\
\end{align*}
identifying the equations above in Eq.~\eqref{a2}, we get the desired result.

%\bibliography{biblio.bib}

\begin{thebibliography}{38}%
\makeatletter
\providecommand \@ifxundefined [1]{%
 \@ifx{#1\undefined}
}%
\providecommand \@ifnum [1]{%
 \ifnum #1\expandafter \@firstoftwo
 \else \expandafter \@secondoftwo
 \fi
}%
\providecommand \@ifx [1]{%
 \ifx #1\expandafter \@firstoftwo
 \else \expandafter \@secondoftwo
 \fi
}%
\providecommand \natexlab [1]{#1}%
\providecommand \enquote  [1]{``#1''}%
\providecommand \bibnamefont  [1]{#1}%
\providecommand \bibfnamefont [1]{#1}%
\providecommand \citenamefont [1]{#1}%
\providecommand \href@noop [0]{\@secondoftwo}%
\providecommand \href [0]{\begingroup \@sanitize@url \@href}%
\providecommand \@href[1]{\@@startlink{#1}\@@href}%
\providecommand \@@href[1]{\endgroup#1\@@endlink}%
\providecommand \@sanitize@url [0]{\catcode `\\12\catcode `\$12\catcode `\&12\catcode `\#12\catcode `\^12\catcode `\_12\catcode `\%12\relax}%
\providecommand \@@startlink[1]{}%
\providecommand \@@endlink[0]{}%
\providecommand \url  [0]{\begingroup\@sanitize@url \@url }%
\providecommand \@url [1]{\endgroup\@href {#1}{\urlprefix }}%
\providecommand \urlprefix  [0]{URL }%
\providecommand \Eprint [0]{\href }%
\providecommand \doibase [0]{https://doi.org/}%
\providecommand \selectlanguage [0]{\@gobble}%
\providecommand \bibinfo  [0]{\@secondoftwo}%
\providecommand \bibfield  [0]{\@secondoftwo}%
\providecommand \translation [1]{[#1]}%
\providecommand \BibitemOpen [0]{}%
\providecommand \bibitemStop [0]{}%
\providecommand \bibitemNoStop [0]{.\EOS\space}%
\providecommand \EOS [0]{\spacefactor3000\relax}%
\providecommand \BibitemShut  [1]{\csname bibitem#1\endcsname}%
\let\auto@bib@innerbib\@empty
%</preamble>
\bibitem [{\citenamefont {Christodoulides}\ \emph {et~al.}(2003)\citenamefont {Christodoulides}, \citenamefont {Lederer},\ and\ \citenamefont {Silberberg}}]{christodoulides2003discretizing}%
  \BibitemOpen
  \bibfield  {author} {\bibinfo {author} {\bibfnamefont {D.~N.}\ \bibnamefont {Christodoulides}}, \bibinfo {author} {\bibfnamefont {F.}~\bibnamefont {Lederer}},\ and\ \bibinfo {author} {\bibfnamefont {Y.}~\bibnamefont {Silberberg}},\ }\bibfield  {title} {\bibinfo {title} {Discretizing light behaviour in linear and nonlinear waveguide lattices},\ }\href {https://doi.org/10.1038/nature01936} {\bibfield  {journal} {\bibinfo  {journal} {Nature}\ }\textbf {\bibinfo {volume} {424}},\ \bibinfo {pages} {817} (\bibinfo {year} {2003})}\BibitemShut {NoStop}%
\bibitem [{\citenamefont {Barral}\ \emph {et~al.}(2020)\citenamefont {Barral}, \citenamefont {Walschaers}, \citenamefont {Bencheikh}, \citenamefont {Parigi}, \citenamefont {Levenson}, \citenamefont {Treps},\ and\ \citenamefont {Belabas}}]{barral2020quantum}%
  \BibitemOpen
  \bibfield  {author} {\bibinfo {author} {\bibfnamefont {D.}~\bibnamefont {Barral}}, \bibinfo {author} {\bibfnamefont {M.}~\bibnamefont {Walschaers}}, \bibinfo {author} {\bibfnamefont {K.}~\bibnamefont {Bencheikh}}, \bibinfo {author} {\bibfnamefont {V.}~\bibnamefont {Parigi}}, \bibinfo {author} {\bibfnamefont {J.~A.}\ \bibnamefont {Levenson}}, \bibinfo {author} {\bibfnamefont {N.}~\bibnamefont {Treps}},\ and\ \bibinfo {author} {\bibfnamefont {N.}~\bibnamefont {Belabas}},\ }\bibfield  {title} {\bibinfo {title} {Quantum state engineering in arrays of nonlinear waveguides},\ }\href {https://doi.org/10.1103/PhysRevA.102.043706} {\bibfield  {journal} {\bibinfo  {journal} {Phys. Rev. A}\ }\textbf {\bibinfo {volume} {102}},\ \bibinfo {pages} {043706} (\bibinfo {year} {2020})}\BibitemShut {NoStop}%
\bibitem [{\citenamefont {Perets}\ \emph {et~al.}(2008)\citenamefont {Perets}, \citenamefont {Lahini}, \citenamefont {Pozzi}, \citenamefont {Sorel}, \citenamefont {Morandotti},\ and\ \citenamefont {Silberberg}}]{perets2008realization}%
  \BibitemOpen
  \bibfield  {author} {\bibinfo {author} {\bibfnamefont {H.~B.}\ \bibnamefont {Perets}}, \bibinfo {author} {\bibfnamefont {Y.}~\bibnamefont {Lahini}}, \bibinfo {author} {\bibfnamefont {F.}~\bibnamefont {Pozzi}}, \bibinfo {author} {\bibfnamefont {M.}~\bibnamefont {Sorel}}, \bibinfo {author} {\bibfnamefont {R.}~\bibnamefont {Morandotti}},\ and\ \bibinfo {author} {\bibfnamefont {Y.}~\bibnamefont {Silberberg}},\ }\bibfield  {title} {\bibinfo {title} {Realization of quantum walks with negligible decoherence in waveguide lattices},\ }\href {https://doi.org/10.1103/PhysRevLett.100.170506} {\bibfield  {journal} {\bibinfo  {journal} {Phys. Rev. Lett.}\ }\textbf {\bibinfo {volume} {100}},\ \bibinfo {pages} {170506} (\bibinfo {year} {2008})}\BibitemShut {NoStop}%
\bibitem [{\citenamefont {Peruzzo}\ \emph {et~al.}(2010)\citenamefont {Peruzzo}, \citenamefont {Lobino}, \citenamefont {Matthews}, \citenamefont {Matsuda}, \citenamefont {Politi}, \citenamefont {Poulios}, \citenamefont {Zhou}, \citenamefont {Lahini}, \citenamefont {Ismail}, \citenamefont {W{\"o}rhoff} \emph {et~al.}}]{peruzzo2010quantum}%
  \BibitemOpen
  \bibfield  {author} {\bibinfo {author} {\bibfnamefont {A.}~\bibnamefont {Peruzzo}}, \bibinfo {author} {\bibfnamefont {M.}~\bibnamefont {Lobino}}, \bibinfo {author} {\bibfnamefont {J.~C.}\ \bibnamefont {Matthews}}, \bibinfo {author} {\bibfnamefont {N.}~\bibnamefont {Matsuda}}, \bibinfo {author} {\bibfnamefont {A.}~\bibnamefont {Politi}}, \bibinfo {author} {\bibfnamefont {K.}~\bibnamefont {Poulios}}, \bibinfo {author} {\bibfnamefont {X.-Q.}\ \bibnamefont {Zhou}}, \bibinfo {author} {\bibfnamefont {Y.}~\bibnamefont {Lahini}}, \bibinfo {author} {\bibfnamefont {N.}~\bibnamefont {Ismail}}, \bibinfo {author} {\bibfnamefont {K.}~\bibnamefont {W{\"o}rhoff}}, \emph {et~al.},\ }\bibfield  {title} {\bibinfo {title} {Quantum walks of correlated photons},\ }\href {https://doi.org/10.1126/science.1193515} {\bibfield  {journal} {\bibinfo  {journal} {Science}\ }\textbf {\bibinfo {volume} {329}},\ \bibinfo {pages} {1500} (\bibinfo {year} {2010})}\BibitemShut {NoStop}%
\bibitem [{\citenamefont {Lahini}\ \emph {et~al.}(2008)\citenamefont {Lahini}, \citenamefont {Avidan}, \citenamefont {Pozzi}, \citenamefont {Sorel}, \citenamefont {Morandotti}, \citenamefont {Christodoulides},\ and\ \citenamefont {Silberberg}}]{lahini2008anderson}%
  \BibitemOpen
  \bibfield  {author} {\bibinfo {author} {\bibfnamefont {Y.}~\bibnamefont {Lahini}}, \bibinfo {author} {\bibfnamefont {A.}~\bibnamefont {Avidan}}, \bibinfo {author} {\bibfnamefont {F.}~\bibnamefont {Pozzi}}, \bibinfo {author} {\bibfnamefont {M.}~\bibnamefont {Sorel}}, \bibinfo {author} {\bibfnamefont {R.}~\bibnamefont {Morandotti}}, \bibinfo {author} {\bibfnamefont {.~f. D.~N.}\ \bibnamefont {Christodoulides}},\ and\ \bibinfo {author} {\bibfnamefont {Y.}~\bibnamefont {Silberberg}},\ }\bibfield  {title} {\bibinfo {title} {Anderson localization and nonlinearity in one-dimensional disordered photonic lattices},\ }\href {https://doi.org/10.1103/PhysRevLett.100.013906} {\bibfield  {journal} {\bibinfo  {journal} {Phys. Rev. Lett.}\ }\textbf {\bibinfo {volume} {100}},\ \bibinfo {pages} {013906} (\bibinfo {year} {2008})}\BibitemShut {NoStop}%
\bibitem [{\citenamefont {Joglekar}\ \emph {et~al.}(2013)\citenamefont {Joglekar}, \citenamefont {Thompson}, \citenamefont {Scott},\ and\ \citenamefont {Vemuri}}]{joglekar2013optical}%
  \BibitemOpen
  \bibfield  {author} {\bibinfo {author} {\bibfnamefont {Y.~N.}\ \bibnamefont {Joglekar}}, \bibinfo {author} {\bibfnamefont {C.}~\bibnamefont {Thompson}}, \bibinfo {author} {\bibfnamefont {D.~D.}\ \bibnamefont {Scott}},\ and\ \bibinfo {author} {\bibfnamefont {G.}~\bibnamefont {Vemuri}},\ }\bibfield  {title} {\bibinfo {title} {Optical waveguide arrays: quantum effects and {PT} symmetry breaking},\ }\href {https://doi.org/10.1051/epjap/2013130240} {\bibfield  {journal} {\bibinfo  {journal} {The European Physical Journal Applied Physics}\ }\textbf {\bibinfo {volume} {63}},\ \bibinfo {pages} {30001} (\bibinfo {year} {2013})}\BibitemShut {NoStop}%
\bibitem [{\citenamefont {Rai}\ \emph {et~al.}(2010)\citenamefont {Rai}, \citenamefont {Das},\ and\ \citenamefont {Agarwal}}]{rai2010quantum}%
  \BibitemOpen
  \bibfield  {author} {\bibinfo {author} {\bibfnamefont {A.}~\bibnamefont {Rai}}, \bibinfo {author} {\bibfnamefont {S.}~\bibnamefont {Das}},\ and\ \bibinfo {author} {\bibfnamefont {G.}~\bibnamefont {Agarwal}},\ }\bibfield  {title} {\bibinfo {title} {Quantum entanglement in coupled lossy waveguides},\ }\href {https://doi.org/10.1364/OE.18.006241} {\bibfield  {journal} {\bibinfo  {journal} {Opt. Express}\ }\textbf {\bibinfo {volume} {18}},\ \bibinfo {pages} {6241} (\bibinfo {year} {2010})}\BibitemShut {NoStop}%
\bibitem [{\citenamefont {Perez-Leija}\ \emph {et~al.}(2016)\citenamefont {Perez-Leija}, \citenamefont {Szameit}, \citenamefont {Ramos-Prieto}, \citenamefont {Moya-Cessa},\ and\ \citenamefont {Christodoulides}}]{perez2016generalized}%
  \BibitemOpen
  \bibfield  {author} {\bibinfo {author} {\bibfnamefont {A.}~\bibnamefont {Perez-Leija}}, \bibinfo {author} {\bibfnamefont {A.}~\bibnamefont {Szameit}}, \bibinfo {author} {\bibfnamefont {I.}~\bibnamefont {Ramos-Prieto}}, \bibinfo {author} {\bibfnamefont {H.}~\bibnamefont {Moya-Cessa}},\ and\ \bibinfo {author} {\bibfnamefont {D.~N.}\ \bibnamefont {Christodoulides}},\ }\bibfield  {title} {\bibinfo {title} {Generalized {S}chr\"odinger cat states and their classical emulation},\ }\href {https://doi.org/10.1103/PhysRevA.93.053815} {\bibfield  {journal} {\bibinfo  {journal} {Phys. Rev. A}\ }\textbf {\bibinfo {volume} {93}},\ \bibinfo {pages} {053815} (\bibinfo {year} {2016})}\BibitemShut {NoStop}%
\bibitem [{\citenamefont {Biggerstaff}\ \emph {et~al.}(2016)\citenamefont {Biggerstaff}, \citenamefont {Heilmann}, \citenamefont {Zecevik}, \citenamefont {Gr{\"a}fe}, \citenamefont {Broome}, \citenamefont {Fedrizzi}, \citenamefont {Nolte}, \citenamefont {Szameit}, \citenamefont {White},\ and\ \citenamefont {Kassal}}]{biggerstaff2016enhancing}%
  \BibitemOpen
  \bibfield  {author} {\bibinfo {author} {\bibfnamefont {D.~N.}\ \bibnamefont {Biggerstaff}}, \bibinfo {author} {\bibfnamefont {R.}~\bibnamefont {Heilmann}}, \bibinfo {author} {\bibfnamefont {A.~A.}\ \bibnamefont {Zecevik}}, \bibinfo {author} {\bibfnamefont {M.}~\bibnamefont {Gr{\"a}fe}}, \bibinfo {author} {\bibfnamefont {M.~A.}\ \bibnamefont {Broome}}, \bibinfo {author} {\bibfnamefont {A.}~\bibnamefont {Fedrizzi}}, \bibinfo {author} {\bibfnamefont {S.}~\bibnamefont {Nolte}}, \bibinfo {author} {\bibfnamefont {A.}~\bibnamefont {Szameit}}, \bibinfo {author} {\bibfnamefont {A.~G.}\ \bibnamefont {White}},\ and\ \bibinfo {author} {\bibfnamefont {I.}~\bibnamefont {Kassal}},\ }\bibfield  {title} {\bibinfo {title} {Enhancing coherent transport in a photonic network using controllable decoherence},\ }\href {https://doi.org/10.1038/ncomms11282} {\bibfield  {journal} {\bibinfo  {journal} {Nat Commun}\ }\textbf {\bibinfo {volume} {7}},\ \bibinfo {pages} {11282} (\bibinfo {year} {2016})}\BibitemShut {NoStop}%
\bibitem [{\citenamefont {Paulisch}\ \emph {et~al.}(2016)\citenamefont {Paulisch}, \citenamefont {Kimble},\ and\ \citenamefont {Gonz{\'a}lez-Tudela}}]{paulisch2016universal}%
  \BibitemOpen
  \bibfield  {author} {\bibinfo {author} {\bibfnamefont {V.}~\bibnamefont {Paulisch}}, \bibinfo {author} {\bibfnamefont {H.}~\bibnamefont {Kimble}},\ and\ \bibinfo {author} {\bibfnamefont {A.}~\bibnamefont {Gonz{\'a}lez-Tudela}},\ }\bibfield  {title} {\bibinfo {title} {Universal quantum computation in waveguide {QED} using decoherence free subspaces},\ }\href {https://doi.org/10.1088/1367-2630/18/4/043041} {\bibfield  {journal} {\bibinfo  {journal} {New J. Phys.}\ }\textbf {\bibinfo {volume} {18}},\ \bibinfo {pages} {043041} (\bibinfo {year} {2016})}\BibitemShut {NoStop}%
\bibitem [{\citenamefont {Perez-Leija}\ \emph {et~al.}(2010)\citenamefont {Perez-Leija}, \citenamefont {Moya-Cessa}, \citenamefont {Szameit},\ and\ \citenamefont {Christodoulides}}]{Perez-Leija:10}%
  \BibitemOpen
  \bibfield  {author} {\bibinfo {author} {\bibfnamefont {A.}~\bibnamefont {Perez-Leija}}, \bibinfo {author} {\bibfnamefont {H.}~\bibnamefont {Moya-Cessa}}, \bibinfo {author} {\bibfnamefont {A.}~\bibnamefont {Szameit}},\ and\ \bibinfo {author} {\bibfnamefont {D.~N.}\ \bibnamefont {Christodoulides}},\ }\bibfield  {title} {\bibinfo {title} {Glauber-{F}ock photonic lattices},\ }\href {https://doi.org/10.1364/OL.35.002409} {\bibfield  {journal} {\bibinfo  {journal} {Opt. Lett.}\ }\textbf {\bibinfo {volume} {35}},\ \bibinfo {pages} {2409} (\bibinfo {year} {2010})}\BibitemShut {NoStop}%
\bibitem [{\citenamefont {Rom\'an-Ancheyta}\ \emph {et~al.}(2017)\citenamefont {Rom\'an-Ancheyta}, \citenamefont {Ramos-Prieto}, \citenamefont {Perez-Leija}, \citenamefont {Busch},\ and\ \citenamefont {Le\'on-Montiel}}]{Ancheyta_2017}%
  \BibitemOpen
  \bibfield  {author} {\bibinfo {author} {\bibfnamefont {R.}~\bibnamefont {Rom\'an-Ancheyta}}, \bibinfo {author} {\bibfnamefont {I.}~\bibnamefont {Ramos-Prieto}}, \bibinfo {author} {\bibfnamefont {A.}~\bibnamefont {Perez-Leija}}, \bibinfo {author} {\bibfnamefont {K.}~\bibnamefont {Busch}},\ and\ \bibinfo {author} {\bibfnamefont {R.~d.~J.}\ \bibnamefont {Le\'on-Montiel}},\ }\bibfield  {title} {\bibinfo {title} {Dynamical {C}asimir effect in stochastic systems: Photon harvesting through noise},\ }\href {https://doi.org/10.1103/PhysRevA.96.032501} {\bibfield  {journal} {\bibinfo  {journal} {Phys. Rev. A}\ }\textbf {\bibinfo {volume} {96}},\ \bibinfo {pages} {032501} (\bibinfo {year} {2017})}\BibitemShut {NoStop}%
\bibitem [{\citenamefont {Keil}\ \emph {et~al.}(2011)\citenamefont {Keil}, \citenamefont {Perez-Leija}, \citenamefont {Dreisow}, \citenamefont {Heinrich}, \citenamefont {Moya-Cessa}, \citenamefont {Nolte}, \citenamefont {Christodoulides},\ and\ \citenamefont {Szameit}}]{PhysRevLett.107.103601}%
  \BibitemOpen
  \bibfield  {author} {\bibinfo {author} {\bibfnamefont {R.}~\bibnamefont {Keil}}, \bibinfo {author} {\bibfnamefont {A.}~\bibnamefont {Perez-Leija}}, \bibinfo {author} {\bibfnamefont {F.}~\bibnamefont {Dreisow}}, \bibinfo {author} {\bibfnamefont {M.}~\bibnamefont {Heinrich}}, \bibinfo {author} {\bibfnamefont {H.}~\bibnamefont {Moya-Cessa}}, \bibinfo {author} {\bibfnamefont {S.}~\bibnamefont {Nolte}}, \bibinfo {author} {\bibfnamefont {D.~N.}\ \bibnamefont {Christodoulides}},\ and\ \bibinfo {author} {\bibfnamefont {A.}~\bibnamefont {Szameit}},\ }\bibfield  {title} {\bibinfo {title} {Classical analogue of displaced {F}ock states and quantum correlations in {G}lauber-{F}ock photonic lattices},\ }\href {https://doi.org/10.1103/PhysRevLett.107.103601} {\bibfield  {journal} {\bibinfo  {journal} {Phys. Rev. Lett.}\ }\textbf {\bibinfo {volume} {107}},\ \bibinfo {pages} {103601} (\bibinfo {year} {2011})}\BibitemShut {NoStop}%
\bibitem [{\citenamefont {Ramos-Prieto}\ \emph {et~al.}(2021)\citenamefont {Ramos-Prieto}, \citenamefont {Uriostegui}, \citenamefont {R\'{e}camier}, \citenamefont {Soto-Eguibar},\ and\ \citenamefont {Moya-Cessa}}]{Ramos_2021}%
  \BibitemOpen
  \bibfield  {author} {\bibinfo {author} {\bibfnamefont {I.}~\bibnamefont {Ramos-Prieto}}, \bibinfo {author} {\bibfnamefont {K.}~\bibnamefont {Uriostegui}}, \bibinfo {author} {\bibfnamefont {J.}~\bibnamefont {R\'{e}camier}}, \bibinfo {author} {\bibfnamefont {F.}~\bibnamefont {Soto-Eguibar}},\ and\ \bibinfo {author} {\bibfnamefont {H.~M.}\ \bibnamefont {Moya-Cessa}},\ }\bibfield  {title} {\bibinfo {title} {Kapitza--{D}irac photonic lattices},\ }\href {https://doi.org/10.1364/OL.437829} {\bibfield  {journal} {\bibinfo  {journal} {Opt. Lett.}\ }\textbf {\bibinfo {volume} {46}},\ \bibinfo {pages} {4690} (\bibinfo {year} {2021})}\BibitemShut {NoStop}%
\bibitem [{\citenamefont {Hern\'{a}ndez-S\'{a}nchez}\ \emph {et~al.}(2023)\citenamefont {Hern\'{a}ndez-S\'{a}nchez}, \citenamefont {Ramos-Prieto}, \citenamefont {Soto-Eguibar},\ and\ \citenamefont {Moya-Cessa}}]{Leonardi_2023}%
  \BibitemOpen
  \bibfield  {author} {\bibinfo {author} {\bibfnamefont {L.}~\bibnamefont {Hern\'{a}ndez-S\'{a}nchez}}, \bibinfo {author} {\bibfnamefont {I.}~\bibnamefont {Ramos-Prieto}}, \bibinfo {author} {\bibfnamefont {F.}~\bibnamefont {Soto-Eguibar}},\ and\ \bibinfo {author} {\bibfnamefont {H.~M.}\ \bibnamefont {Moya-Cessa}},\ }\bibfield  {title} {\bibinfo {title} {Exact solution for the interaction of two decaying quantized fields},\ }\href {https://doi.org/10.1364/OL.503837} {\bibfield  {journal} {\bibinfo  {journal} {Opt. Lett.}\ }\textbf {\bibinfo {volume} {48}},\ \bibinfo {pages} {5435} (\bibinfo {year} {2023})}\BibitemShut {NoStop}%
\bibitem [{\citenamefont {Urz\'{u}a}\ \emph {et~al.}(2024)\citenamefont {Urz\'{u}a}, \citenamefont {Ramos-Prieto},\ and\ \citenamefont {Moya-Cessa}}]{Urzua_2024}%
  \BibitemOpen
  \bibfield  {author} {\bibinfo {author} {\bibfnamefont {A.~R.}\ \bibnamefont {Urz\'{u}a}}, \bibinfo {author} {\bibfnamefont {I.}~\bibnamefont {Ramos-Prieto}},\ and\ \bibinfo {author} {\bibfnamefont {H.~M.}\ \bibnamefont {Moya-Cessa}},\ }\bibfield  {title} {\bibinfo {title} {Integrated optical wave analyzer using the discrete fractional {F}ourier transform},\ }\href {https://doi.org/10.1364/JOSAB.533919} {\bibfield  {journal} {\bibinfo  {journal} {J. Opt. Soc. Am. B}\ }\textbf {\bibinfo {volume} {41}},\ \bibinfo {pages} {2358} (\bibinfo {year} {2024})}\BibitemShut {NoStop}%
\bibitem [{\citenamefont {Perez-Leija}\ \emph {et~al.}(2011)\citenamefont {Perez-Leija}, \citenamefont {Moya-Cessa}, \citenamefont {Soto-Eguibar}, \citenamefont {Aguilar-Loreto},\ and\ \citenamefont {Christodoulides}}]{perez2011}%
  \BibitemOpen
  \bibfield  {author} {\bibinfo {author} {\bibfnamefont {A.}~\bibnamefont {Perez-Leija}}, \bibinfo {author} {\bibfnamefont {H.}~\bibnamefont {Moya-Cessa}}, \bibinfo {author} {\bibfnamefont {F.}~\bibnamefont {Soto-Eguibar}}, \bibinfo {author} {\bibfnamefont {O.}~\bibnamefont {Aguilar-Loreto}},\ and\ \bibinfo {author} {\bibfnamefont {D.~N.}\ \bibnamefont {Christodoulides}},\ }\bibfield  {title} {\bibinfo {title} {Classical analogues to quantum nonlinear coherent states in photonic lattices},\ }\href {https://doi.org/10.1016/j.optcom.2010.12.005} {\bibfield  {journal} {\bibinfo  {journal} {Opt. Commun.}\ }\textbf {\bibinfo {volume} {284}},\ \bibinfo {pages} {1833} (\bibinfo {year} {2011})}\BibitemShut {NoStop}%
\bibitem [{\citenamefont {Sch{\"a}fer}\ \emph {et~al.}(2020)\citenamefont {Sch{\"a}fer}, \citenamefont {Fukuhara}, \citenamefont {Sugawa}, \citenamefont {Takasu},\ and\ \citenamefont {Takahashi}}]{schafer2020tools}%
  \BibitemOpen
  \bibfield  {author} {\bibinfo {author} {\bibfnamefont {F.}~\bibnamefont {Sch{\"a}fer}}, \bibinfo {author} {\bibfnamefont {T.}~\bibnamefont {Fukuhara}}, \bibinfo {author} {\bibfnamefont {S.}~\bibnamefont {Sugawa}}, \bibinfo {author} {\bibfnamefont {Y.}~\bibnamefont {Takasu}},\ and\ \bibinfo {author} {\bibfnamefont {Y.}~\bibnamefont {Takahashi}},\ }\bibfield  {title} {\bibinfo {title} {Tools for quantum simulation with ultracold atoms in optical lattices},\ }\href {https://doi.org/10.1038/s42254-020-0195-3} {\bibfield  {journal} {\bibinfo  {journal} {Nat. Rev. Phys.}\ }\textbf {\bibinfo {volume} {2}},\ \bibinfo {pages} {411} (\bibinfo {year} {2020})}\BibitemShut {NoStop}%
\bibitem [{\citenamefont {Vicencio}\ \emph {et~al.}(2015)\citenamefont {Vicencio}, \citenamefont {Cantillano}, \citenamefont {Morales-Inostroza}, \citenamefont {Real}, \citenamefont {Mej\'{\i}a-Cort\'es}, \citenamefont {Weimann}, \citenamefont {Szameit},\ and\ \citenamefont {Molina}}]{vicencio2015observation}%
  \BibitemOpen
  \bibfield  {author} {\bibinfo {author} {\bibfnamefont {R.~A.}\ \bibnamefont {Vicencio}}, \bibinfo {author} {\bibfnamefont {C.}~\bibnamefont {Cantillano}}, \bibinfo {author} {\bibfnamefont {L.}~\bibnamefont {Morales-Inostroza}}, \bibinfo {author} {\bibfnamefont {B.}~\bibnamefont {Real}}, \bibinfo {author} {\bibfnamefont {C.}~\bibnamefont {Mej\'{\i}a-Cort\'es}}, \bibinfo {author} {\bibfnamefont {S.}~\bibnamefont {Weimann}}, \bibinfo {author} {\bibfnamefont {A.}~\bibnamefont {Szameit}},\ and\ \bibinfo {author} {\bibfnamefont {M.~I.}\ \bibnamefont {Molina}},\ }\bibfield  {title} {\bibinfo {title} {Observation of localized states in lieb photonic lattices},\ }\href {https://doi.org/10.1103/PhysRevLett.114.245503} {\bibfield  {journal} {\bibinfo  {journal} {Phys. Rev. Lett.}\ }\textbf {\bibinfo {volume} {114}},\ \bibinfo {pages} {245503} (\bibinfo {year} {2015})}\BibitemShut {NoStop}%
\bibitem [{\citenamefont {Meany}\ \emph {et~al.}(2015)\citenamefont {Meany}, \citenamefont {Gr{\"a}fe}, \citenamefont {Heilmann}, \citenamefont {Perez-Leija}, \citenamefont {Gross}, \citenamefont {Steel}, \citenamefont {Withford},\ and\ \citenamefont {Szameit}}]{meany2015laser}%
  \BibitemOpen
  \bibfield  {author} {\bibinfo {author} {\bibfnamefont {T.}~\bibnamefont {Meany}}, \bibinfo {author} {\bibfnamefont {M.}~\bibnamefont {Gr{\"a}fe}}, \bibinfo {author} {\bibfnamefont {R.}~\bibnamefont {Heilmann}}, \bibinfo {author} {\bibfnamefont {A.}~\bibnamefont {Perez-Leija}}, \bibinfo {author} {\bibfnamefont {S.}~\bibnamefont {Gross}}, \bibinfo {author} {\bibfnamefont {M.~J.}\ \bibnamefont {Steel}}, \bibinfo {author} {\bibfnamefont {M.~J.}\ \bibnamefont {Withford}},\ and\ \bibinfo {author} {\bibfnamefont {A.}~\bibnamefont {Szameit}},\ }\bibfield  {title} {\bibinfo {title} {Laser written circuits for quantum photonics},\ }\href {https://doi.org/https://doi.org/10.1002/lpor.201500061} {\bibfield  {journal} {\bibinfo  {journal} {Laser \& Photonics Reviews}\ }\textbf {\bibinfo {volume} {9}},\ \bibinfo {pages} {363} (\bibinfo {year} {2015})}\BibitemShut {NoStop}%
\bibitem [{\citenamefont {Efremidis}\ and\ \citenamefont {Christodoulides}(2002)}]{efremidis2002discrete}%
  \BibitemOpen
  \bibfield  {author} {\bibinfo {author} {\bibfnamefont {N.~K.}\ \bibnamefont {Efremidis}}\ and\ \bibinfo {author} {\bibfnamefont {D.~N.}\ \bibnamefont {Christodoulides}},\ }\bibfield  {title} {\bibinfo {title} {Discrete solitons in nonlinear zigzag optical waveguide arrays with tailored diffraction properties},\ }\href {https://doi.org/10.1103/PhysRevE.65.056607} {\bibfield  {journal} {\bibinfo  {journal} {Phys. Rev. E}\ }\textbf {\bibinfo {volume} {65}},\ \bibinfo {pages} {056607} (\bibinfo {year} {2002})}\BibitemShut {NoStop}%
\bibitem [{\citenamefont {Szameit}\ \emph {et~al.}(2008)\citenamefont {Szameit}, \citenamefont {Pertsch}, \citenamefont {Nolte}, \citenamefont {T{\"u}nnermann},\ and\ \citenamefont {Lederer}}]{szameit2008long}%
  \BibitemOpen
  \bibfield  {author} {\bibinfo {author} {\bibfnamefont {A.}~\bibnamefont {Szameit}}, \bibinfo {author} {\bibfnamefont {T.}~\bibnamefont {Pertsch}}, \bibinfo {author} {\bibfnamefont {S.}~\bibnamefont {Nolte}}, \bibinfo {author} {\bibfnamefont {A.}~\bibnamefont {T{\"u}nnermann}},\ and\ \bibinfo {author} {\bibfnamefont {F.}~\bibnamefont {Lederer}},\ }\bibfield  {title} {\bibinfo {title} {Long-range interaction in waveguide lattices},\ }\href {https://doi.org/10.1103/PhysRevA.77.043804} {\bibfield  {journal} {\bibinfo  {journal} {Phys. Rev. A}\ }\textbf {\bibinfo {volume} {77}},\ \bibinfo {pages} {043804} (\bibinfo {year} {2008})}\BibitemShut {NoStop}%
\bibitem [{\citenamefont {Villegas-Mart{\'\i}nez}\ \emph {et~al.}(2022)\citenamefont {Villegas-Mart{\'\i}nez}, \citenamefont {Moya-Cessa},\ and\ \citenamefont {Soto-Eguibar}}]{villegas2022modeling}%
  \BibitemOpen
  \bibfield  {author} {\bibinfo {author} {\bibfnamefont {B.}~\bibnamefont {Villegas-Mart{\'\i}nez}}, \bibinfo {author} {\bibfnamefont {H.}~\bibnamefont {Moya-Cessa}},\ and\ \bibinfo {author} {\bibfnamefont {F.}~\bibnamefont {Soto-Eguibar}},\ }\bibfield  {title} {\bibinfo {title} {Modeling displaced squeezed number states in waveguide arrays},\ }\href {https://doi.org/https://doi.org/10.1016/j.physa.2022.128265} {\bibfield  {journal} {\bibinfo  {journal} {Physica A: Statistical Mechanics and its applications}\ }\textbf {\bibinfo {volume} {608}},\ \bibinfo {pages} {128265} (\bibinfo {year} {2022})}\BibitemShut {NoStop}%
\bibitem [{\citenamefont {Wang}\ \emph {et~al.}(2010)\citenamefont {Wang}, \citenamefont {Huang},\ and\ \citenamefont {Yu}}]{wang2010nontrivial}%
  \BibitemOpen
  \bibfield  {author} {\bibinfo {author} {\bibfnamefont {G.}~\bibnamefont {Wang}}, \bibinfo {author} {\bibfnamefont {J.~P.}\ \bibnamefont {Huang}},\ and\ \bibinfo {author} {\bibfnamefont {K.~W.}\ \bibnamefont {Yu}},\ }\bibfield  {title} {\bibinfo {title} {Nontrivial {B}loch oscillations in waveguide arrays with second-order coupling},\ }\href {https://doi.org/10.1364/OL.35.001908} {\bibfield  {journal} {\bibinfo  {journal} {Opt. Lett.}\ }\textbf {\bibinfo {volume} {35}},\ \bibinfo {pages} {1908} (\bibinfo {year} {2010})}\BibitemShut {NoStop}%
\bibitem [{\citenamefont {Dreisow}\ \emph {et~al.}(2011)\citenamefont {Dreisow}, \citenamefont {Wang}, \citenamefont {Heinrich}, \citenamefont {Keil}, \citenamefont {T{\"u}nnermann}, \citenamefont {Nolte},\ and\ \citenamefont {Szameit}}]{dreisow2011observation}%
  \BibitemOpen
  \bibfield  {author} {\bibinfo {author} {\bibfnamefont {F.}~\bibnamefont {Dreisow}}, \bibinfo {author} {\bibfnamefont {G.}~\bibnamefont {Wang}}, \bibinfo {author} {\bibfnamefont {M.}~\bibnamefont {Heinrich}}, \bibinfo {author} {\bibfnamefont {R.}~\bibnamefont {Keil}}, \bibinfo {author} {\bibfnamefont {A.}~\bibnamefont {T{\"u}nnermann}}, \bibinfo {author} {\bibfnamefont {S.}~\bibnamefont {Nolte}},\ and\ \bibinfo {author} {\bibfnamefont {A.}~\bibnamefont {Szameit}},\ }\bibfield  {title} {\bibinfo {title} {Observation of anharmonic bloch oscillations},\ }\href {https://doi.org/10.1364/OL.36.003963} {\bibfield  {journal} {\bibinfo  {journal} {Opt. Lett.}\ }\textbf {\bibinfo {volume} {36}},\ \bibinfo {pages} {3963} (\bibinfo {year} {2011})}\BibitemShut {NoStop}%
\bibitem [{\citenamefont {Qi}\ \emph {et~al.}(2014)\citenamefont {Qi}, \citenamefont {Feng}, \citenamefont {Wang}, \citenamefont {Xu}, \citenamefont {Zhu},\ and\ \citenamefont {Zheng}}]{qi2014photon}%
  \BibitemOpen
  \bibfield  {author} {\bibinfo {author} {\bibfnamefont {F.}~\bibnamefont {Qi}}, \bibinfo {author} {\bibfnamefont {Z.}~\bibnamefont {Feng}}, \bibinfo {author} {\bibfnamefont {Y.}~\bibnamefont {Wang}}, \bibinfo {author} {\bibfnamefont {P.}~\bibnamefont {Xu}}, \bibinfo {author} {\bibfnamefont {S.}~\bibnamefont {Zhu}},\ and\ \bibinfo {author} {\bibfnamefont {W.}~\bibnamefont {Zheng}},\ }\bibfield  {title} {\bibinfo {title} {Photon-number correlations in waveguide lattices with second order coupling},\ }\href {https://doi.org/10.1088/2040-8978/16/12/125007} {\bibfield  {journal} {\bibinfo  {journal} {J. Opt.}\ }\textbf {\bibinfo {volume} {16}},\ \bibinfo {pages} {125007} (\bibinfo {year} {2014})}\BibitemShut {NoStop}%
\bibitem [{\citenamefont {Longhi}(2006)}]{longhi2006tunneling}%
  \BibitemOpen
  \bibfield  {author} {\bibinfo {author} {\bibfnamefont {S.}~\bibnamefont {Longhi}},\ }\bibfield  {title} {\bibinfo {title} {Tunneling escape in optical waveguide arrays with a boundary defect},\ }\href {https://doi.org/10.1103/PhysRevE.74.026602} {\bibfield  {journal} {\bibinfo  {journal} {Phys. Rev. E}\ }\textbf {\bibinfo {volume} {74}},\ \bibinfo {pages} {026602} (\bibinfo {year} {2006})}\BibitemShut {NoStop}%
\bibitem [{\citenamefont {Istrate}\ and\ \citenamefont {Sargent}(2005)}]{istrate2005photonic}%
  \BibitemOpen
  \bibfield  {author} {\bibinfo {author} {\bibfnamefont {E.}~\bibnamefont {Istrate}}\ and\ \bibinfo {author} {\bibfnamefont {E.~H.}\ \bibnamefont {Sargent}},\ }\bibfield  {title} {\bibinfo {title} {Photonic crystal waveguide analysis using interface boundary conditions},\ }\href {https://doi.org/10.1109/JQE.2004.841615} {\bibfield  {journal} {\bibinfo  {journal} {IEEE journal of quantum electronics}\ }\textbf {\bibinfo {volume} {41}},\ \bibinfo {pages} {461} (\bibinfo {year} {2005})}\BibitemShut {NoStop}%
\bibitem [{\citenamefont {Ablowitz}\ and\ \citenamefont {Musslimani}(2003)}]{ablowitz2003discrete}%
  \BibitemOpen
  \bibfield  {author} {\bibinfo {author} {\bibfnamefont {M.~J.}\ \bibnamefont {Ablowitz}}\ and\ \bibinfo {author} {\bibfnamefont {Z.~H.}\ \bibnamefont {Musslimani}},\ }\bibfield  {title} {\bibinfo {title} {Discrete spatial solitons in a diffraction-managed nonlinear waveguide array: a unified approach},\ }\href {https://doi.org/https://doi.org/10.1016/S0167-2789(03)00226-4} {\bibfield  {journal} {\bibinfo  {journal} {Physica D: Nonlinear Phenomena}\ }\textbf {\bibinfo {volume} {184}},\ \bibinfo {pages} {276} (\bibinfo {year} {2003})}\BibitemShut {NoStop}%
\bibitem [{\citenamefont {Trompeter}\ \emph {et~al.}(2003)\citenamefont {Trompeter}, \citenamefont {Peschel}, \citenamefont {Pertsch}, \citenamefont {Lederer}, \citenamefont {Streppel}, \citenamefont {Michaelis},\ and\ \citenamefont {Br{\"a}uer}}]{trompeter2003tailoring}%
  \BibitemOpen
  \bibfield  {author} {\bibinfo {author} {\bibfnamefont {H.}~\bibnamefont {Trompeter}}, \bibinfo {author} {\bibfnamefont {U.}~\bibnamefont {Peschel}}, \bibinfo {author} {\bibfnamefont {T.}~\bibnamefont {Pertsch}}, \bibinfo {author} {\bibfnamefont {F.}~\bibnamefont {Lederer}}, \bibinfo {author} {\bibfnamefont {U.}~\bibnamefont {Streppel}}, \bibinfo {author} {\bibfnamefont {D.}~\bibnamefont {Michaelis}},\ and\ \bibinfo {author} {\bibfnamefont {A.}~\bibnamefont {Br{\"a}uer}},\ }\bibfield  {title} {\bibinfo {title} {Tailoring guided modes in waveguide arrays},\ }\href {https://doi.org/10.1364/OE.11.003404} {\bibfield  {journal} {\bibinfo  {journal} {Opt. Express}\ }\textbf {\bibinfo {volume} {11}},\ \bibinfo {pages} {3404} (\bibinfo {year} {2003})}\BibitemShut {NoStop}%
\bibitem [{\citenamefont {London}(1926)}]{london1926jacobischen}%
  \BibitemOpen
  \bibfield  {author} {\bibinfo {author} {\bibfnamefont {F.}~\bibnamefont {London}},\ }\bibfield  {title} {\bibinfo {title} {{\"U}ber die jacobischen transformationen der quantenmechanik},\ }\href {https://doi.org/10.1007/BF01397484} {\bibfield  {journal} {\bibinfo  {journal} {Z. Physik}\ }\textbf {\bibinfo {volume} {37}},\ \bibinfo {pages} {915} (\bibinfo {year} {1926})}\BibitemShut {NoStop}%
\bibitem [{\citenamefont {Susskind}\ and\ \citenamefont {Glogower}(1964)}]{susskind1964quantum}%
  \BibitemOpen
  \bibfield  {author} {\bibinfo {author} {\bibfnamefont {L.}~\bibnamefont {Susskind}}\ and\ \bibinfo {author} {\bibfnamefont {J.}~\bibnamefont {Glogower}},\ }\bibfield  {title} {\bibinfo {title} {Quantum mechanical phase and time operator},\ }\href {https://doi.org/10.1103/PhysicsPhysiqueFizika.1.49} {\bibfield  {journal} {\bibinfo  {journal} {Physics Physique Fizika}\ }\textbf {\bibinfo {volume} {1}},\ \bibinfo {pages} {49} (\bibinfo {year} {1964})}\BibitemShut {NoStop}%
\bibitem [{\citenamefont {Arfken}\ \emph {et~al.}(2011)\citenamefont {Arfken}, \citenamefont {Weber},\ and\ \citenamefont {Harris}}]{arfken2011mathematical}%
  \BibitemOpen
  \bibfield  {author} {\bibinfo {author} {\bibfnamefont {G.}~\bibnamefont {Arfken}}, \bibinfo {author} {\bibfnamefont {H.}~\bibnamefont {Weber}},\ and\ \bibinfo {author} {\bibfnamefont {F.}~\bibnamefont {Harris}},\ }\href {https://books.google.com.mx/books?id=JOpHkJF-qcwC} {\emph {\bibinfo {title} {Mathematical Methods for Physicists: A Comprehensive Guide}}}\ (\bibinfo  {publisher} {Elsevier Science},\ \bibinfo {year} {2011})\BibitemShut {NoStop}%
\bibitem [{\citenamefont {Le\'on-Montiel}\ and\ \citenamefont {Moya-Cessa}(2011)}]{doi:10.1142/S0219749911007319}%
  \BibitemOpen
  \bibfield  {author} {\bibinfo {author} {\bibfnamefont {R.~d.~J.}\ \bibnamefont {Le\'on-Montiel}}\ and\ \bibinfo {author} {\bibfnamefont {H.}~\bibnamefont {Moya-Cessa}},\ }\bibfield  {title} {\bibinfo {title} {Modeling non-linear coherent states in fiber array},\ }\href {https://doi.org/10.1142/S0219749911007319} {\bibfield  {journal} {\bibinfo  {journal} {International Journal of Quantum Information}\ }\textbf {\bibinfo {volume} {09}},\ \bibinfo {pages} {349} (\bibinfo {year} {2011})}\BibitemShut {NoStop}%
\bibitem [{\citenamefont {Makris}\ and\ \citenamefont {Christodoulides}(2006)}]{makris2006method}%
  \BibitemOpen
  \bibfield  {author} {\bibinfo {author} {\bibfnamefont {K.~G.}\ \bibnamefont {Makris}}\ and\ \bibinfo {author} {\bibfnamefont {D.~N.}\ \bibnamefont {Christodoulides}},\ }\bibfield  {title} {\bibinfo {title} {Method of images in optical discrete systems},\ }\href {https://doi.org/10.1103/PhysRevE.73.036616} {\bibfield  {journal} {\bibinfo  {journal} {Phys. Rev. E}\ }\textbf {\bibinfo {volume} {73}},\ \bibinfo {pages} {036616} (\bibinfo {year} {2006})}\BibitemShut {NoStop}%
\bibitem [{\citenamefont {Dattoli}\ \emph {et~al.}(1991)\citenamefont {Dattoli}, \citenamefont {Torre}, \citenamefont {Lorenzutta}, \citenamefont {Maino},\ and\ \citenamefont {Chiccoli}}]{dattoli1991theory}%
  \BibitemOpen
  \bibfield  {author} {\bibinfo {author} {\bibfnamefont {G.}~\bibnamefont {Dattoli}}, \bibinfo {author} {\bibfnamefont {A.}~\bibnamefont {Torre}}, \bibinfo {author} {\bibfnamefont {S.}~\bibnamefont {Lorenzutta}}, \bibinfo {author} {\bibfnamefont {G.}~\bibnamefont {Maino}},\ and\ \bibinfo {author} {\bibfnamefont {C.}~\bibnamefont {Chiccoli}},\ }\bibfield  {title} {\bibinfo {title} {Theory of generalized {B}essel functions.-ii},\ }\href {https://doi.org/10.1007/BF02723125} {\bibfield  {journal} {\bibinfo  {journal} {Il Nuovo Cimento B (1971-1996)}\ }\textbf {\bibinfo {volume} {106}},\ \bibinfo {pages} {21} (\bibinfo {year} {1991})}\BibitemShut {NoStop}%
\bibitem [{\citenamefont {Szameit}\ \emph {et~al.}(2007)\citenamefont {Szameit}, \citenamefont {Dreisow}, \citenamefont {Pertsch}, \citenamefont {Nolte},\ and\ \citenamefont {T{\"u}nnermann}}]{szameit2007control}%
  \BibitemOpen
  \bibfield  {author} {\bibinfo {author} {\bibfnamefont {A.}~\bibnamefont {Szameit}}, \bibinfo {author} {\bibfnamefont {F.}~\bibnamefont {Dreisow}}, \bibinfo {author} {\bibfnamefont {T.}~\bibnamefont {Pertsch}}, \bibinfo {author} {\bibfnamefont {S.}~\bibnamefont {Nolte}},\ and\ \bibinfo {author} {\bibfnamefont {A.}~\bibnamefont {T{\"u}nnermann}},\ }\bibfield  {title} {\bibinfo {title} {Control of directional evanescent coupling in fs laser written waveguides},\ }\href {https://doi.org/10.1364/OE.15.001579} {\bibfield  {journal} {\bibinfo  {journal} {Opt. Express}\ }\textbf {\bibinfo {volume} {15}},\ \bibinfo {pages} {1579} (\bibinfo {year} {2007})}\BibitemShut {NoStop}%
\bibitem [{\citenamefont {Szameit}\ \emph {et~al.}(2006)\citenamefont {Szameit}, \citenamefont {Bl{\"o}mer}, \citenamefont {Burghoff}, \citenamefont {Pertsch}, \citenamefont {Nolte},\ and\ \citenamefont {T{\"u}nnermann}}]{szameit2006hexagonal}%
  \BibitemOpen
  \bibfield  {author} {\bibinfo {author} {\bibfnamefont {A.}~\bibnamefont {Szameit}}, \bibinfo {author} {\bibfnamefont {D.}~\bibnamefont {Bl{\"o}mer}}, \bibinfo {author} {\bibfnamefont {J.}~\bibnamefont {Burghoff}}, \bibinfo {author} {\bibfnamefont {T.}~\bibnamefont {Pertsch}}, \bibinfo {author} {\bibfnamefont {S.}~\bibnamefont {Nolte}},\ and\ \bibinfo {author} {\bibfnamefont {A.}~\bibnamefont {T{\"u}nnermann}},\ }\bibfield  {title} {\bibinfo {title} {Hexagonal waveguide arrays written with fs-laser pulses},\ }\href {https://doi.org/https://doi.org/10.1007/s00340-005-2127-4} {\bibfield  {journal} {\bibinfo  {journal} {Appl. Phys. B}\ }\textbf {\bibinfo {volume} {82}},\ \bibinfo {pages} {507} (\bibinfo {year} {2006})}\BibitemShut {NoStop}%
\end{thebibliography}
\end{document}